\documentclass[a4paper,12pt]{article}

\usepackage{amstext,amsmath,amssymb,amsthm}  

\usepackage[normalem]{ulem}  
\usepackage{graphicx}  
\usepackage{bm} 

\usepackage[comma,sort,authoryear]{natbib}
\usepackage{hypernat}  

\usepackage{vmargin}  
\usepackage{authblk}

\usepackage{xcolor}
\definecolor{dark-red}{rgb}{0.8,0.15,0.15}
\definecolor{dark-blue}{rgb}{0.15,0.15,0.6}
\definecolor{medium-blue}{rgb}{0,0,0.8}

\usepackage[pdftex,breaklinks,colorlinks]{hyperref}
\hypersetup{  
  pdftitle =
    {Some lessons for us scientists (too) from the ``Sokal affair''},
  linkcolor={dark-red},
  citecolor={dark-blue},
  urlcolor={medium-blue}
}

\begin{document}

\numberwithin{equation}{section}
\numberwithin{figure}{section}
\allowdisplaybreaks[1]  

\noindent
{\LARGE Some lessons for us scientists (too) from the ``Sokal affair''}\\[12pt]
{\large Pablo Echenique-Robba}\\[2pt]
{\small \emph{Instituto de Qu\'{\i}mica F\'{\i}sica Rocasolano, CSIC, Spain}}\\
{\small \emph{BIFI, ZCAM, DFTUZ, University of Zaragoza, Spain}}\\[2pt]
{\small \href{mailto:pablo.echenique.robba@gmail.com}
             {\texttt{pablo.echenique.robba@gmail.com}} }\\[-2pt]
{\small \href{http://www.pabloecheniquerobba.com}
             {\texttt{http://www.pabloecheniquerobba.com}} } \\[10pt]
\today

\vspace{10pt}
\begin{abstract}
In this little non-technical piece, I argue that some of the lessons that can
be learnt from the bold action carried out in 1996 by the physicist Alan Sokal
and typically known as the ``Sokal affair'' not only apply to some sector of
the humanities (which was the original target of the hoax), but also (with much
less intensity, but still) to the hardest sciences.
\end{abstract}
\vspace{10pt}

The reader probably knows about the famous ``Sokal affair''. This refers to an
illuminating action designed and carried out by Alan Sokal in 1996. The physics
professor at NYU submitted an article entitled ``Transgressing the boundaries:
Towards a transformative hermeneutics of quantum gravity'' to \emph{Social
Text}, a high-impact, well known academic journal of postmodern cultural
studies. In Sokal's own words, what he wanted to test was this: ``Would a
leading North American journal of cultural studies ---whose editorial
collective includes such luminaries as Fredric Jameson and Andrew Ross---
publish an article liberally salted with nonsense if (a) it sounded good and
(b) it flattered the editors' ideological preconceptions?'' \citep{Sokal1996}.
That is, he deliberately sent and absurd article which was ``a pastiche of
left-wing cant, fawning references, grandiose quotations, and outright
nonsense\ldots structured around the silliest quotations [by postmodernist
academics] he could find about mathematics and physics'' \citep{Wikipedia2013},
an article that he wrote ``so that any competent physicist or mathematician (or
undergraduate physics or math major) would realize that it is a spoof''
\citep{Sokal1996}.

The answer to Sokal's question was (unfortunately for our trust in the
collective intelligence of humankind) \emph{yes}. The article got published in
\emph{Social Text} \citep{Sokal1996a}, and he soon denounced it was a hoax in
the journal \emph{Lingua Franca} \citep{Sokal1996}.

The whole business is very interesting and several considerations enter the
mix:

First, it is important to remark that Sokal is a declared ``leftist'' (whatever
this 1-dimensional classification of political tendencies may mean in these
times), and one of his objectives was to denounce the anti-scientific,
anti-rationalistic attitude of a large part of the left. This is important, it
is also very sad (specially for rationalistic ``leftists''), it is as valid now
as it was in 1996, but I will not focus on it here.

Another lesson that the Sokal affair teaches us is that believing in things
that make us feel good can be dangerous (to say the least). This is explicit in
his second point, ``(b) it flattered the editors' ideological preconceptions''.
Why? Because, for most people, confirming preconceptions feels good and
contradicting them feels bad.

The lesson is in fact more general than this, since confirming preconceptions
is by no means the only way of producing nice warm feelings out of beliefs and
intellectual conclusions. The sources are multiple: believing that there is
life after death, believing that medicines (such as homeopathy) exist with no
secondary effects and capable of curing virtually everything, believing that
you are right about a point and most people is wrong (Neil Armstrong didn't go
the Moon), and many more, all make people feel good for obvious reasons.
Another way of putting it is due to David Albert. In a great interview in which
he tries to control the damage of having been inadvertently talked into
participating in the shameful film ``What the bleep do we know?'', he explains
that the main difference between the views which science helps us to arrive to
and those defended by the Vatican or by the producers and fans of the film is
that the second are (and must be!) ``therapeutic'', while the views suggested
by science do not have to be (and typically are not) \citep{Albert2012}.
Science forces us to be honest to ourselves (when it works well), and this
includes not letting warm feelings lead us to ``therapeutic'' but false
conclusions about the world.

Of course, these blatantly obvious concessions to one's feelings are nowhere to
be found among successful scientists in the hard sciences, but I think that
something more subtle and related to this \emph{is} in place. No serious
scientist will let herself be influenced by not wanting to die, or by the
desire of having a cure-it-all medicine; that is too childish. But it is also
clear that some pressure exists to arrive to conclusions that, say, confirm
what was said in previous publications by the same scientist, that are
consistent with the achievements that were promised in the last funding grant,
or that do not go too much against the usual way of understanding things in
the corresponding field (thus making the peer-review process ``smoother'').
Depending on the personality of the scientist, these pressures will be enough
to lead the discourse to wrong (but convenient) conclusions\ldots or not.
After all, confirming and thus increasing the importance of one's past
results, getting nice grants, and not having to struggle too much with 
referees suspicious of our heterodoxy \emph{does} feel good. And scientists 
are human ---despite many opinions on the contrary.

A very nice example is one that \cite{Dennett2009} likes to use. It seems that 
when ``The origin of species'' was published a Robert Beverley MacKenzie 
answered Darwin with a long criticism containing the following paragraph:

\begin{quote}
{\small But in the Theory with which we have to deal, Absolute Ignorance is
the artificer, so that we may enunciate as the fundamental principle of the
whole system, that in order TO MAKE A PERFECT AND BEAUTIFUL MACHINE IT IS NOT
REQUISITE TO KNOW HOW TO MAKE IT [capital (outraged) letters in the original].
This proposition will be found, on a careful examination, to express in a
condensed form the essential purport of the Theory, and to express in a few
words all Mr Darwin's meaning; who, by a strange inversion of reasoning, seems
to think Absolute Ignorance fully qualified to take the place of Absolute
Wisdom in all the achievements of creative skill.}
\end{quote}

As Dennett says, ``Exactly!'' This piece of text is one of the most accurate,
distilled and insightful descriptions of what Darwin had achieved, thus proving
that MacKenzie was a clever fellow who had read the whole treatise and who had
understood it thoroughly. However, he not only disagreed, he hated Darwin's
conclusion. Why? Because it went against one of the beliefs that he held
dearest and which made him good and warm inside: that an intelligent creator
was behind life in general and humans in particular.

When doing science, it is convenient to remember Feynman's famous aphorism:
``The first principle is that you must not fool yourself, and you are the 
easiest person to fool'' \citep[chap.~10]{Feynman1999} ---which is, of course,
also applicable to me.

An example much closer to my line of work pertains some analyses of hybrid
quantum-classical models in chemical physics. I will not go so far as to state
that the authors of the corresponding papers are guilty of ``fooling
themselves'' with respect to their quantitative conclusions (after all, the
conclusions tend to be numerically validated, and rigorously so). But I cannot
help realizing the uncritical way in which some ill-defined and even false
statements are repeated and (it seems) carried forward from one introduction to
the next. For example, in the otherwise excellent review by \citep{Truhlar2007}
(and by no means \emph{only} there) we can find the statement that Ehrenfest
evolution is unitary ---which, being non-linear, is obviously not
\citep{Alonso2011,Alonso2012c}. I think that this should make us a bit
suspicious about the hypotheses from which these papers start, and maybe also
about the interpretation of the quantitative results. Of course, the same
caution should be exercised if we catch \emph{ourselves} repeating something
uncritically. Nobody is free from making this kind of mistakes.

A third lesson that we can learn from Sokal's hoax is emphasized in the book he
later wrote together with Jean Bricmont \citep{Sokal1998}, namely, that
postmodernist writers like to misuse scientific and mathematical concepts to
support their ``arguments'' (e.g., a given postmodernist argued that the
famous equation $E=mc^2$ is a ``sexed equation'' because ``it privileges the
speed of light over other speeds that are vitally necessary to us'').

Again, this is an extreme (and funny!) case of a much more general practice. It
is common that all kinds of thinkers use concepts from a ``more fundamental''
(or just different) field to sound more clever and attach more weight to their
arguments. The trick is very simple in its workings: Since you write mostly
for your colleagues (who work in the same field as you), it is very likely that
they do not understand very well the borrowed concepts that you are planning to
use. However, they are not certain that you don't understand them either (hey,
maybe you spent your last sabbatical reading about formal logic, who knows).
Hence, if you use the concepts with gravity and (apparent) self-assurance, they
might think that you know what you mean, and (not knowing formal logic) they
will probably assent silently. Try it, it works!

As I say, this is a common pitfall in scientific discourses and it is not
always so obvious and ridiculous as in postmodernist papers. Normally, the
discipline from which the borrowed concepts come from is very close to the one
in which the author is an expert, thus making the \emph{bona fide} assumption
that she knows what she means more reasonable. Also, since the borrowed
concepts \emph{are} in fact close, the author might misuse them, but only
slightly. She is not an expert, but she is not completely alien to them
either.

I claim here that theoretical physicists (including myself) are sometimes
guilty of this kind of slight misuses related to philosophical, mathematical or
biological concepts; mathematicians borrow gaily from physics; biologists from
physics and chemistry; and theoretical chemists from quantum physics and
mathematics.

Finally, in my opinion probably the greatest warning coming from the Sokal
affair is related to the dangers of using ambiguous and vague language. One of
the points that \cite{Sokal1998} discuss in their book is indeed ``manipulating
words and phrases that are, in fact, meaningless'' or the use of ``deliberately
obscure language'', but my content is that this is not again something
circumscribed to the most absurd postmodernist texts only. This is a practice
which is all-pervading; and not only in science, but in society as a whole. It
fact, it is in science where the greatest efforts have been made to sharpen the
language, to be precise, to deal with unique meanings, to disambiguate natural
words, and I think that this is one of the main reasons behind the enormous
achievements of our scientific and technological society (the scientific
method: yes; the aforementioned honest approach to nature: yes; the precise
language: no doubt, too).

You see, if a word has three (or twenty!) possible meanings and we do not
start by declaring with care and precision which one of them we are thinking
about, it is very likely that I am using one of the meanings and you are using
a different one. If the discourse contains not only one such word but many of
them, the odds that we do not understand each other are very high. We will
very probably end talking past each other or, in the best of cases, we will
strongly disagree and we will be amazed how the other person can possibly hold
such absurd beliefs about the world. If we also include the possibility that
some of the words' meanings have blurred boundaries (bald, tall, teenager),
that some words have no meaning at all (chakra, aura, karma, luck), or we
accept composed concepts made of words that have meaning independently but it
is destroyed upon combination (quantum healing, negative vibrations), then you
can imagine how bad the situation can get.

Many conversations are like this in everyday life and, unfortunately, also in
science (as I say, to a much lower degree, but still). Even in quantum
mechanics, one of the finest theories ever created by us humans, many
conceptual problems have survived for almost a century very likely due (in
part) to the use of ambiguous language in its very axiomatic foundations
\citep{Bricmont2013a,Echenique-Robba2013}. In this case, the word ``measure''
seems to be the likely culprit.

It takes a lot of work to try to be as precise as possible in every sentence,
in every word, but I think it is worth the effort. I think it is better to
write less, to publish less, but to think deeper. To stop and ask ourselves
from time to time: ``What do I really mean with `wordX'? Am I sure that I am
using it properly? Am I sure that I can define it sharply and neatly?'' I think
that being extremely careful with the meaning of words is not just being picky
and wasting others' time, but it can serve to prove that some widely accepted
hypotheses are wrong, and to arrive to new and applicable results.

The lessons of the ``Sokal affair'' do not apply to cultural studies only, but
also to science.

\phantomsection
\addcontentsline{toc}{section}{References}

\end{document}